# Wind- and Operation-Induced Vibration Measurements of the Main Reflector of the Nobeyama 45 m Radio Telescope

Ikumi Hashimoto[1], Masakatsu Chiba[1,*], Nozomi Okada[2], Hideo Ogawa[2], Ryohei Kawabe[3,4],

Tetsuhiro Minamidani[5], Yoichi Tamura[6], Kimihiro Kimura[6]

[1]Dept. Aerospace Engineering, Graduate School of Engineering, Osaka Prefecture University, 1-1 Gakuen-cho, Sakai, Osaka 599-8531, Japan

[2]Dept. Physical Science, Graduate School of Science, Osaka Prefecture University, 1-1 Gakuen-cho, Sakai, Osaka 599-8531, Japan

[3]Div. Science, National Astronomical Observatory of Japan, 2-21-1 Osawa, Mitaka, Tokyo 181-8588, Japan,

[4]The Graduate University for Advanced Studies (SOKENDAI), 2-21-1 Osawa, Mitaka, Tokyo 181-0015, Japan

[5]National Astronomical Observatory of Japan Nobeyama, 462-2, Minamisaku, Nobeyama, Nagano 384-1305, Japan

[6]Div. Particle and Astrophysical Science, Graduate School of Science, Nagoya University, Furo-cho, Chikusa-ku, Nagoya, Aichi 464-8602, Japan

\* Corresponding author:
E-mail address: chiba@aero.osakafu-u.ac.jp (*M. Chiba*)

**Abstract**

**Purpose** As deformations of the main reflector of a radio telescope directly affect the observations, the evaluation of the deformation is extremely important. Dynamic characteristics of the main reflector of the Nobeyama 45 m radio telescope, Japan, are measured under two conditions: the first is when the pointing observation is in operation, and the second is when the reflector is stationary and is subjected to wind loads when the observation is out of operation.

**Methods** Dynamic characteristics of the main reflector are measured using piezoelectric accelerometers.

**Results and Conclusion** When the telescope is in operation, a vibration mode with one nodal line horizontally or vertically on the reflector are induced, depending on whether the reflector is moving in the azimuthal or elevational planes, whereas under windy conditions, vibration modes that have two to four nodal lines are simultaneously induced. The predominant mode is dependent on the direction of wind loads.

## 1. Introduction

Radio telescopes can observe interstellar gases that do not radiate visible light. Observation is conducted by condensing radio waves from a distant celestial body using a large parabolic main reflector. As high resolution is needed to obtain a clear image, the diameter of a radio telescope can be up to 10 times that of an optical telescope. In general, the resolution of the telescope is proportional to the diameter and inversely proportional to the observed wavelengths. As the wavelength of radio waves is long compared to other electromagnetic waves and approximately 10,000 times that of visible light, a high resolution can be obtained by increasing the diameter size.

For an ideal shape, the error derived from the parabolic shape of the main reflector must be less than 1/12th the observed wavelength, and the deformation of the main reflector should be minimized as much as possible when under various external disturbances (e.g., gravity, heat, and wind loads). In addition, deformation of the main reflector during observation causes a loss of reflector efficiency. Evaluation of the deformation of the main reflector is critical as it may cause a reduction in the pointing accuracy and reduced aperture efficiency.

One of the causes of vibration in the main reflector is the driving of the telescope during the observation. For the Subaru Coronagraphic Extreme Adaptive Optics (SCExAO) instrument currently being developed for the Subaru Telescope, Lozi at al. [1] identified the cause of low-frequency vibrations when the telescope is driven in azimuth and elevation. In addition, they proposed a Linear Quadratic Gaussian controller to control the vibrations. Concerning the dynamic characteristics of the telescope, Hugh et al. [2] measured vibration forces using various types of observatory equipment as well as the transmission of these forces using a shaker and accelerometer. McBride et al. [3] measured the vibration characteristics of the Daniel K. Inouye Solar Telescope (DKIST) using an inertia-mass shaker and accelerometers.

Regarding the potential external disturbances, we next discuss the wind load. For a general structure under a wind load, Davenport's equation [4], in which the time history data of wind loads is employed, can be used. Studies have been conducted on the dynamic deformation of a radio telescope main reflector. Some of these are summarized as follows.

Gawronski et al. [5] developed a wind force model and simulated wind-induced vibrations as well as pointing errors of the Deep Space Network (DSN) antenna, in particular the Direct Signal Suppression (DSS)-13 antenna with 34 m diameter. Ukita et al. [6] measured the response of the 10 m ASTE telescope in Chile under a wind load using accelerometers. They found that the wind caused low-frequency vibrations throughout the main reflector, which induced torsional movement around the yoke arm. This and the pitching motion were major contributors to the pointing errors. In addition, they clarified that for motion in the azimuthal plane, frontal-direction wind had a greater effect than did side- or tail-direction winds. For the Giant Magellan Telescope (GMT) in Chile, which has a 25 m diameter reflector, Irarrazaval et al. [7] proposed a finite element model (FEM) for the telescope structure and a computational fluid dynamics (CFD) model for wind loads. Zhang et al. [8] analyzed the impact of the flexible oscillation of a reflective surface resulting from a stochastic wind disturbance on pointing accuracy using structural deformations derived from dynamic equations based on modal analysis. Liu et al. [9] conducted wind tunnel test and CFD simulation to investigate wind loads and response characteristics of a 1/200 scale model of the 110m reflector which will be built in Xinjiang. For reference, as a similar structure of a telescope reflector, Paetzold et al. [10] conducted vibration test and CFD simulation for a parabolic trough solar collector. Andre et al. [11] also conducted CFD simulation of a similar collector.

Studies on vibrational reduction of telescopes due to wind loads have also been conducted. Glaese et al. [12] proposed a tuned mass damper as a passive approach and a reaction mass actuator for an active approach for the GMT, and Du et al. [13] proposed an active damper system based on a linear motor actuator.

Regular studies have been conducted on the Nobeyama Radio Observatory (NRO) 45 m radio telescope. In April 1999, Smith et al. [14] measured dynamic pointing variations of the telescope in both the azimuth and elevation directions and measured the error in the frequency as 0.9 Hz in the azimuth twist mode of the main reflector, 1.6 Hz in the elevation mode of the main reflector, and 3.0 Hz in the secondary reflector mode. In October 1999, Blough et al. [15] and Smith et al. [16] conducted measurements using 44 pieces of piezoelectric accelerometers of two types: PCB3701 and PCB U393C. They found some vibrational mode frequencies during low-speed wind (i.e., less than 2 m/s) with frequencies of 0.9, 1.8, 2.1, and 2.6–3 Hz. Under high-speed wind (i.e., 6–8 m/s), they measured higher mode responses but did not predict the vibrational modes. Ukita [17] measured the deformations of the NRO telescope under wind loads using four LED lamps on the main reflector and two CCD cameras on the central hub.

In this study, the dynamic characteristics of a main reflector of the NRO 45 m Radio Telescope were measured using piezoelectric accelerometers under two conditions, namely, during the observation and when the reflector was subjected to wind loads while the telescope was not in operation. From the measurements, not only the frequencies of the responses of the main reflector but also the vibration modes of the responses were determined for the first time.

## 2. Nobeyama 45 m Radio Telescope

The Nobeyama 45 m radio telescope is found in Nobeyama, Nagano, Japan [18]. It is the largest radio telescope that can observe millimeter wavelengths, with the observable frequency band as wide as 20–150 GHz. The specifications and the figures are in Table 1 and Fig. 1, respectively.

As the diameter of the reflector is 45 m, it is not possible to ignore the deformation of the reflector due to its weight, which depends on the elevation position of the reflector. It is possible to employ the truss structure in the reflector structure to minimize induced deformation as much as possible. In addition, for the design of the reflector frame structure, a homologous deformation method is employed, which makes it possible to achieve an optimum shape even if the direction of the weight changes when the elevation position changes. However, even under these considerations, the influence of the strong wind load, which deteriorates observation precision, cannot be ignored.

*2.2. Meteorological station*

The wind direction, wind velocity and atmospheric temperature were recorded every minute at the meteorological station 75 m north of the NRO, which is 30 m in height (see Fig. 2).

*2.3. Definitions of angles of reflector*

The pointing direction of the reflector is defined by the azimuth angle, $Az$, and elevation angle, $El$, as shown in Fig. 3. For example, when the reflector moves from a horizontal direction ($El = 0$), to the zenith ($El = 90°$), $Az = 0$, which corresponds to the clockwise direction, as shown in Fig. 4.

## 3. Measuring system

*3.1. Piezoelectric accelerometers*

At the beginning of this study, four kinds of piezoelectric accelerometer were selected. After a preliminary experiment, two types of accelerometers were chosen- PCB 393B12 and SHOWA SOKKI 2472, as shown in Fig. 5 and Table 2. The selection was based on the minimum measurable frequency value. These accelerometers have a sensitivity up to 100 times greater than that of those used for general structure measurement. PCB 393B12 weighs 210 g and SHOWA SOKKI 2472 weighs 50 g. PCB 393B12 was used in the measurement of ASTE telescope by Smith at al. [14].

Displacement data was obtained by twice integrating the acceleration data before cutting the noise using bypass filter.

*3.2. Measuring devices set-up*

*3.2.1. Locations of piezoelectric accelerometers*

Six accelerometers were fixed on the reflector: four at 90° intervals around the edge (hereafter referred to as "up", "right", "down", and "left"), one at 315° clockwise from "up", and one in the center, as seen in Fig. 6(a).

Fig. 6(b) depicts the inner structure of the reflector. Each accelerometer was set on an aluminum plate, which was clamped to a truss beam using four M8 bolts through a Teflon sheet. A plastic plate was added for insulation. This system was then mounted perpendicular to the reflector surface, as seen in Fig. 6(c). Fig. 6(d) represent the PCB accelerometer set on the reflector in the left position. SHOWA SOKKI accelerometers were used in five other positions.

*3.2.2. Locations of thermocouples*

To measure the temperature at the accelerometers, thermocouples were attached to the upper and center positions of the reflector.

*3.2.3. Configuration of measuring devises*

A block diagram depicting the measuring devices is shown in Fig. 7, and the measuring devices' connected data logger (GL900) is shown in Table A1 of the Appendix, and the details of the measurement instruments are shown in Table A2 of the Appendix.

The acceleration signals were input to Ch. 1- Ch. 6 of the data logger, and the thermocouple signals were input to Ch. 7 and Ch. 8. The data logger and the signal conditioner were set in the optical guide room at the center of the reflector, then the accelerometers and the signal conditioner were connected with low noise cables of 14–71 m long. In addition, the signal conditioner and PC, which were in the equipment room 80 m from the optical room, were connected by an 80 m long LAN cable.

# 4. Measured results

Measurements were conducted under two scenarios: the first was when the pointing observation was in operation, and the second was when the reflector was subjected to wind loads when the observation was out of operation. In the first scenario, the impact of the telescope turning on and off was investigated, while in the second scenario, we investigated the influence of wind on the reflector response.

## *4.1 Measurement during pointing observation*

### *4.1.1. Pointing observation*

During pointing observation, while driving the telescope, the direction of the reflector was scanned in the azimuth and elevation directions, alternately. Our method for following the celestial body is described in the four steps below (see Fig. 8(a)). The numbers below correspond to the positions in Fig. 8(b). Variations of the azimuth angle, $Az$, and the elevation angle, $El$, are shown in Fig. 9 as an example.

  Step 1: Observe **off point** except the targeted celestial body: ①
  Step 2: Scan three points in the azimuth direction: ②-④
  Step 3: Observe **off point**: ⑤
  Step 4: Scan three points in the elevation direction: ⑥-⑧.

### *4.1.2. Under weak wind conditions*

*Measurement date: 2018/12/23 (Sun) 1:51–1:53, average wind speed: W=0.5 m/s, atmospheric temperature: 3.3°C*

#### *4.1.2.1. Scanning in the azimuth direction*

The accelerations recorded are shown in Fig. 10, when the scanning was conducted in the azimuthal direction with nominal variation in the elevation direction. In the figure, six acceleration signals are presented with the azimuth encoder signal. ①~⑤ correspond to the pointing position as shown in Fig. 8(b). From the figure, one can determine that the large accelerations coincide with the velocity change of the telescope, even if the angle of change is very small.

Precise analysis was conducted on the framed region in Fig. 10 and detailed acceleration signals are shown in Fig. 11(a). The signal from the right accelerometer is shown in a red line, while that from the left is shown with a blue line. They are out of phase with each other, which indicates that the reflector is vibrating in an asymmetric mode, as shown in Fig. 11(b), i.e., with one vertical nodal line of vibration mode with $N=1$. Displacements of the two accelerometers are shown in Fig. 11(c), from which one can clearly see the out of phase motions in the left and the right parts of the reflector. The maximum displacement is 0.38 mm. The FFT analysis shown in Fig. 11(c) shows that the predominant frequency is 1.56 Hz.

Large acceleration occurs when the driving velocity of the telescope changes, as is shown above. It is stipulated that the observation should start five minutes after the velocity change during the actual operation of observation. The acceleration signals, displacement and FFT analysis five minutes after the velocity change are shown in Figs. 12(a), (b), and (c), respectively. The displacement becomes small ($30.8\ \mu m$), and there is no predominant frequency in the FFT analysis, which improves the stipulation.

#### *4.1.2.2. Scanning in the elevation direction*

The results shown in Fig. 13 were recorded when the scanning was conducted in the elevation direction with nearly no variation in the azimuth direction. In Fig. 13(a), six acceleration signals are presented with the elevation encoder signal. Fig. 13(c) shows the displacements from the upper and lower accelerometers from which one can clearly see the out of phase motions in the upper and lower parts of the reflector as shown in Fig. 13(b). With one horizontal nodal line, $N=1$, the maximum displacement is 0.66 mm and five minutes later, it is $49.3\ \mu m$. The FFT analysis shown in Fig. 13(d) indicates that the predominant frequency is 1.76 Hz.

*4.1.3. Under strong wind*

*Measurement date: 2018/12/02 (Sun) 9:00–12:00, average wind speed: W=5.0 m/s, wind direction: from back side, atmospheric temperature: 7.6°C*

The above results were measured when the wind velocity was as low as $W$=0.5 m/s. In the measurement taken while scanning in the azimuth direction, another vibration mode of the reflector was observed under strong wind conditions. In this case, wind blew towards the back side of the reflector (Fig. 14) with average wind speed $W$=5.0 m/s, and elevation angle $El \approx 36°$. Although scanning was conducted in the azimuth direction, contrary to the previous case, the azimuth angle did not change. Therefore, the influence of the scanning seemed to be small.

Six acceleration signals are presented with the azimuth encoder signal in Fig. 15(a), when the azimuth angle is constant. Separating out this data, one can calculate the accelerations of the right and left, and up and down sensors, as shown in Fig. 15(b). Accelerations of the right and the left, and the up and the down sensors are in-phase, while those of the right-left and the up-down are out of phase. From this, the vibration mode can be assumed, as in Fig. 15(c), with number of nodal lines $N$=2. The FFT analysis shown in Fig. 15(d) indicates that the predominant frequency is 2.74 Hz. The maximum deflection is $187\ \mu m$, as shown in Fig. 15(e), which is approximately six times as large as the maximum deflection when the wind speed is low. The deflection exceeds the stipulated critical value of $100\ \mu m$.

In the above example, when the telescope was scanning, the influence of the wind was the predominant factor. In the following section, we describe the influence of the wind when the telescope was stationary.

*4.2. Measurement under strong wind*

In the second scenario, during the pointing observation, the influence of wind direction on the reflector vibration was investigated. The elevation angle was maintained at $El = 20°$, while the azimuth angle was changed in $45°$ intervals.

*4.2.1. Measuring condition*

Wind velocity was relatively high ($W \approx$ 1.0–10.7 m/s), and wind direction was 126°–257°. Measurements were conducted while keeping the reflector in standby mode for approximately five minutes.

*Measurement date: 2019/1/15 (Wed) 17:00–18:00, $W \approx 1.0$–10.7 m/s, wind direction: 126°–257°, atmospheric temperature 1.5°C*

*4.2.2. Measured results*

The FFT analyses for four cases of the wind direction are presented in Fig. 16. Wind direction in relation to the reflector are (a) diagonal front: $W \approx 5.7$ m/s, (b) side: $W \approx 7.9$ m/s, (c) diagonal back: $W \approx 5.7$ m/s, and (d) back side: $W \approx 13.3$ m/s. From these figures, one can see that the reflector vibrates simultaneously with multiple natural frequencies (i.e., 2.6 Hz, 3.1 Hz, 4.1 Hz), and the predominant vibration mode depends on the wind direction in relation to the reflector.

Here, we shall concentrate on case (d) when the highest wind speed of $W \approx 13.3$ m/s occurred, as an example. Acceleration waves are shown in Fig. 17(a), and the enlarged view of the framed part of Fig. 17(a) is shown in Fig. 17(b). Responses of the right, the left, the upper and the lower positions are all in phase, except the 45° position, which suggests that the vibration mode has four nodal lines $N$=4, as shown in Fig. 17(c). The FFT analysis shown in Fig. 16(d) indicates that this mode is 4.10 Hz, and the acceleration is 20 mG, which is about 0.4 mm in amplitude.

### 4.3. Measured natural frequencies and vibration modes

The observed natural frequencies and vibration modes are summarized in Table 3 alongside the measured conditions, in which the frequency value in bold is the predominant one under each condition. The observed natural frequencies are plotted with the number of nodal lines, $N$, of the vibration mode in Fig. 18 with orange circles. From the figure one can see that as $N$ increases, the natural frequencies increase. Furthermore, the vibration mode corresponds to the predominant frequency shown in Fig. 16(b), which is presumed to be the $N$=3 mode, which can be validated by fitting the frequency variation curve with $N$, as shown in Fig. 18.

In Fig. 18, the modified natural frequency values for a flat circular plate with free edge are plotted with blue circles, and those for a hemi-spherical shell with angle $\alpha = 90°$ (Fig. 19(a)), and a spherical shell with angle $\alpha = 60°$ (Fig. 19(b)) are plotted with yellow and gray circles, respectively. These are vibration modes where the number of nodal circles is zero. Each value is equated with those of the measured $N$=2 value. Details of the frequencies of a circular plate and a spherical shell are in Appendix II. Fig. 18 compares the measured results with the theoretical values for these three geometries.

## 5. Conclusions

The dynamic characteristics of the main reflector of the Nobeyama 45 m radio telescope were measured using six piezoelectric accelerometers under two conditions: the first was when the pointing observation was in operation, and the second was the reflector was subjected to wind loads while out of operation.

When the pointing observation was in operation, the natural vibration with $N$=1 mode was induced. The frequency was 1.56 Hz when the telescope was driving in the azimuth direction, and 1.76 Hz when it was driving in elevation direction.

While the telescope was stationary and under high wind speed conditions, multiple natural vibration modes were simultaneously induced on the reflector with frequencies 2.62–2.74 Hz, 3.11–3.22 Hz, 4.10–4.11 Hz, which corresponded to $N$=2, 3, 4 modes, respectively. The predominant vibration mode was dependent on the direction of the wind acting on the reflector.

The maximum deflection of the reflector under pointing observation was 0.38 mm in the azimuth direction and 0.66 mm in the elevation direction, under low wind speeds (i.e., 0.5 m/s). Just after the telescope was driven, and during the observation, the maximum deflections were 30.8 μm and 49.3 μm for the azimuth and elevation directions, respectively. This had no influence on the operation of the observation. However, under strong wind conditions (i.e., 5.0 m/s) the maximum deflection became approximately six times that of the maximum deflection under low wind conditions, which caused the pointing error.

**Acknowledgment**

The authors thank Mr. T. Ueda of Nagoya University and Mr. Y. Fukasaku of Tsukuba University for preparation of the measurement devices. The authors are also grateful to Messrs. T. Kanzawa, K. Handa, T, Wada, T. Kurakami, Nobeyama Radio Observatory, for help with conducting the measurements. Part of this work was supported by JSPS KAKENHI (Grant No. 17H06206) and NAOJ Research Coordination Committee, NINS.

Appendix I

The measured data that was input into data logger GL 900 is summarized in Table A1, while the data from the measuring devices is summarized in Table A2.

Appendix II Natural Frequencies of a circular plate and a spherical shell

The natural circular frequency, $\omega$, of a circular plate with free edge condition is given by eq. (A1), where $R$ is radius, $D$ is bending rigidity, $\rho$ is density, $h$ is thickness, and $\nu$ is Poisson ratio, see Colwell at al. [19]. Parameter $\lambda$ is shown in Table A3.

$$\omega = \frac{\lambda^2}{R^2}\sqrt{\frac{D}{\rho h}} \qquad (A1)$$

The natural circular frequency of a spherical shell is given by eq. (A2), where $E$ is Young's modulus by Niordson [20]. Parameter $c$ is shown in Table A4, and is dependent on $\alpha$, which is shown in Fig. 19, and on $h/R$. Here, we assumed $h/R$=0.04.

$$\omega = c\frac{h}{R^2}\sqrt{\frac{G}{\rho}}, \quad G = \frac{E}{2(1+\nu)} \qquad (A2)$$

Table 1 Specifications of Nobeyama 45 m Radio Telescope (NAOJ Nobeyama [18]).

| | |
|---|---|
| Diameter of antenna | 45 m |
| Surface accuracy | 0.1 mm |
| Observation frequency | 20~150 GHz |
| Angular resolution | 0.004° |
| Mass | 700 tons |

Table 2 Specifications of accelerometers.

| | ① PCB | ② SHOWA SOKKI |
|---|---|---|
| Model | 393B12 | 2472 |
| Measurement range [G] | ±0.5 | ±0.5 |
| Sensitivity [V/G] | 10 | 10 |
| Frequency range | 0.15Hz～1 kHz | 0.2Hz～1.3 kHz |
| Spectral noise $\mu G / \sqrt{Hz}$ | 1.3 (1 Hz) | 0.28 (2 Hz) |
| Temperature [°C] | －45~82 | 0~80 |
| Mass [g] | 210 | 50 |
| Diameter [mm] | 33.2 | 25.1 |
| Price ¥ | 213,000 | 128,000 |
| Remark | Used on ASTE. | |

Table 3 Observed frequencies and vibration modes of the main-reflector: Hz, bold: significant frequency.

| Reflector condition | Wind velocity $W$ m/s | Wind direction | $N=1$ | $N=2$ | $N=3$ | N=4 | Maximum deflection [mm] | Deflection under observation [μm] | Factor |
|---|---|---|---|---|---|---|---|---|---|
| Pointing observation. Speed change | 0.5 | | 1.56 | | | | 0.38 | 30.8 | Az-driving |
| | | | 1.76 | | | | 0.66 | 49.3 | El-driving |
| Pointing observation. Constant speed | 5.0 | Back | | 2.74 | | | | 187 | (Az-driving) +wind |
| Stationary | 5.7 | Front | | **2.64** | 3.13 | | | | wind |
| Stationary | 7.9 | Side | | | **3.13** | | | | wind |
| Stationary | 5.7 | Oblique back | | 2.62 | **3.11** | 4.11 | | | wind |
| Stationary | 13.3 | Back | | 2.64 | 3.22 | **4.10** | 0.4 | | wind |
| Vibration mode | | | | | | | | | |

Table A1　Input channels of data logger : GL900.

| channel | | Position (cable length) | Accelerometer Model | S.N. | Sensitivity[V/g] |
|---|---|---|---|---|---|
| 1 | Accelerometer | Right (50 m) | 2472 | 1206 | 10.1 |
| 2 | Accelerometer | Left (45 m) | 393B12 | | 10.73 |
| 3 | Accelerometer | Up (26 m) | 2472 | 1208 | 10.1 |
| 4 | Accelerometer | 315° (35 m) | 2472 | 1207 | 9.8 |
| 5 | Accelerometer | Down (71 m) | 2472 | 1209 | 10.2 |
| 6 | Accelerometer | Center (14 m) | 2472 | 1210 | 9.8 |
| 7 | Thermocouple | Up (26 m) | | | |
| 8 | Thermocouple | Down (14 m) | | | |

Table A2　Details of measuring devices.

| | | Model |
|---|---|---|
| Signal conditioner | PCB | 482C05 (4ch) |
| Data logger | GRAPHTEC | GL900－8, Input：Voltage |
| Accelerometer | SHOWASOKKI | 2472 |
| Accelerometer | PCB | 393B12 |
| Thermocouple | KOKKA ELECTRIC | TC-K0.2FEP |
| PC | Computer KOBO | STYLE-15FH053-i7-HNSS |
| BNC coaxial cable | MISUMI | PP-RG58AU |

Table A3　Values of $\lambda^2$ for a completely free circular plate : $v$=0.33, $m$=0 [19].

| $N$ | $\lambda^2$ |
|---|---|
| 2 | 5.253 |
| 3 | 12.23 |
| 4 | 21.6* |
| 5 | 33.1* |

*Values true within 2 percent.

Table A4　Values of $c$ for an open spherical shell: $v$=0.3 [20].

| α | $N$ | $h/R$ | | | |
| | | 0.04 | 0.02 | 0.01 | 0.005 |
|---|---|---|---|---|---|
| 60° | 2 | 3.085 | 3.145 | 3.184 | 3.210 |
| | 3 | 7.647 | 7.929 | 8.132 | 8.267 |
| | 4 | 13.659 | 14.350 | 14.901 | 15.284 |
| 90° | 2 | 2.007 | 2.052 | 2.081 | 2.100 |
| | 3 | 5.356 | 5.577 | 5.724 | 5.819 |
| | 4 | 9.834 | 10.411 | 10.817 | 11.086 |

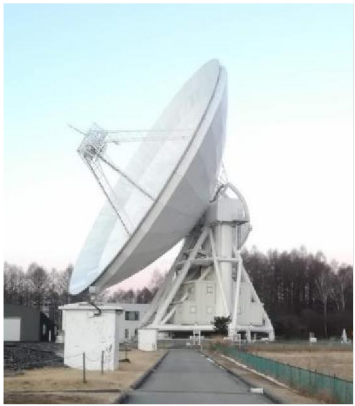 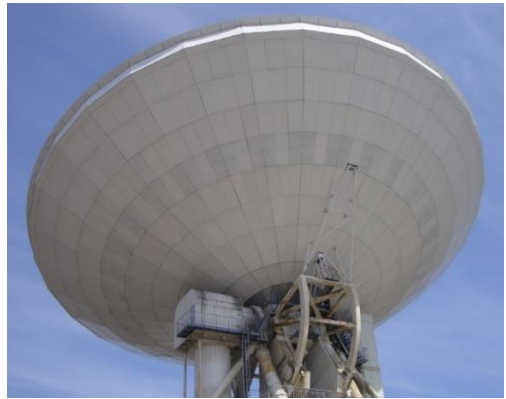

(a) Side view.  (b) Back structures.

Fig. 1. Nobeyama 45 m Radio Telescope.

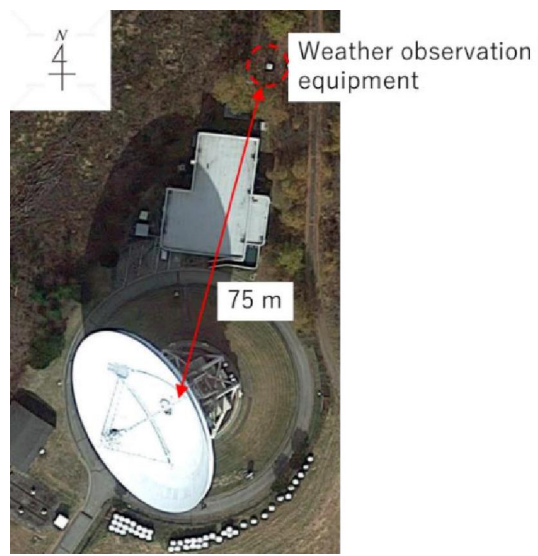

Fig. 2. The weather observation equipment.

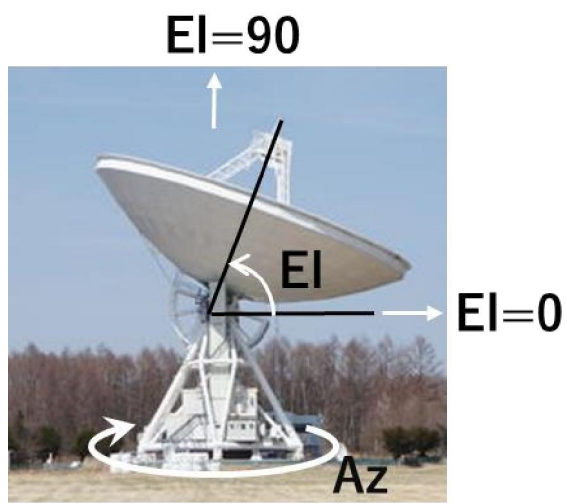

Fig. 3. Definitions of Azimuth angle $Az$ and Elevation angle $El$.

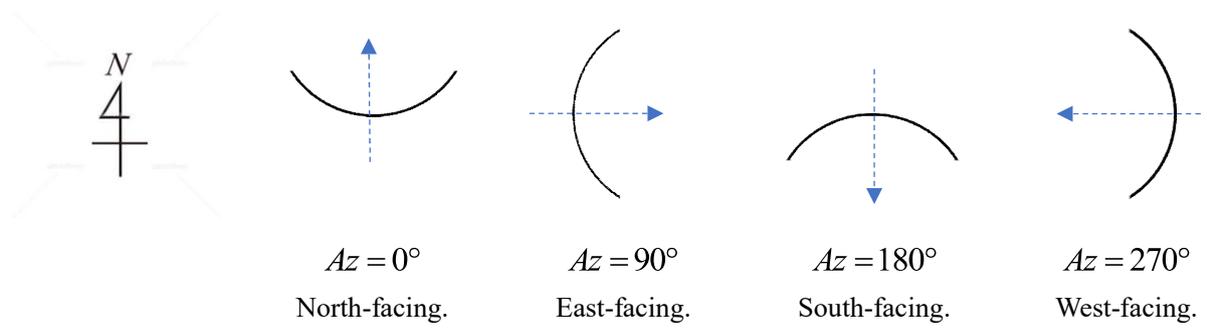

Fig. 4. Azimuth directions: *Az*.

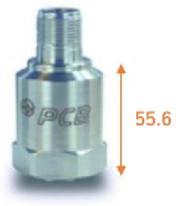 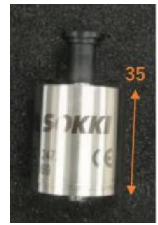

(a) ① PCB 393B12   (b) ② SHOWA SOKKI 2472

Fig. 5.　Accelerometers, mm.

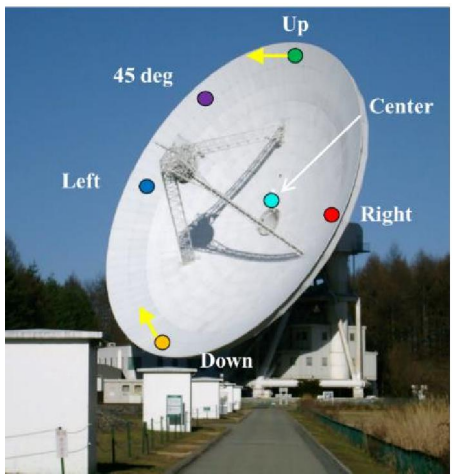

(a) Positions of six accelerometers.

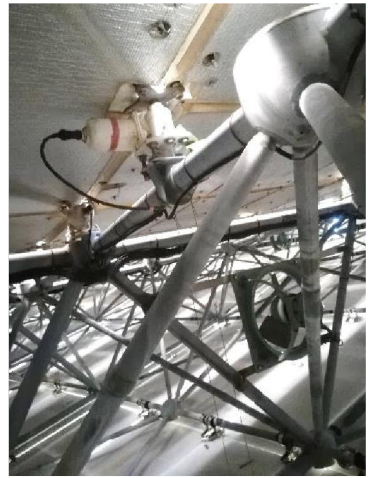

(b) Internal structure.

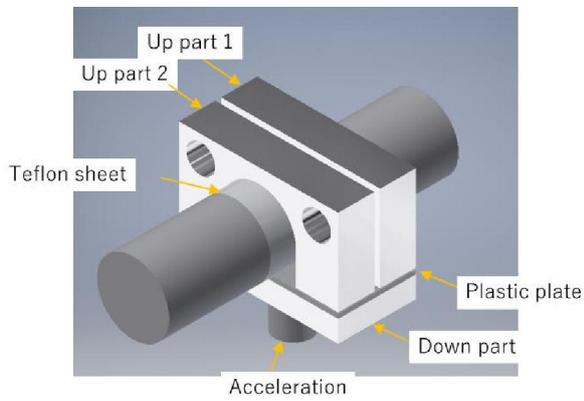

(c) Accelerometer mounting jig; Aluminum, mass: 715 g)

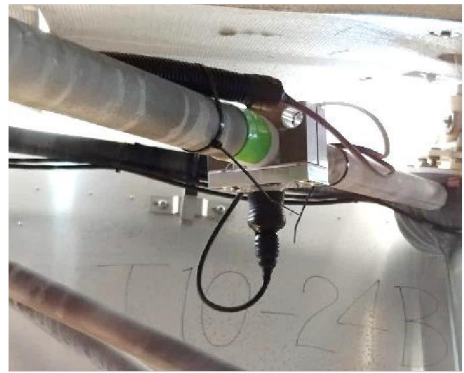

(d) Accelerometer on *Left* (393B12).

Fig. 6.   Positions of accelerometers on the Nobeyama 45 m telescope.

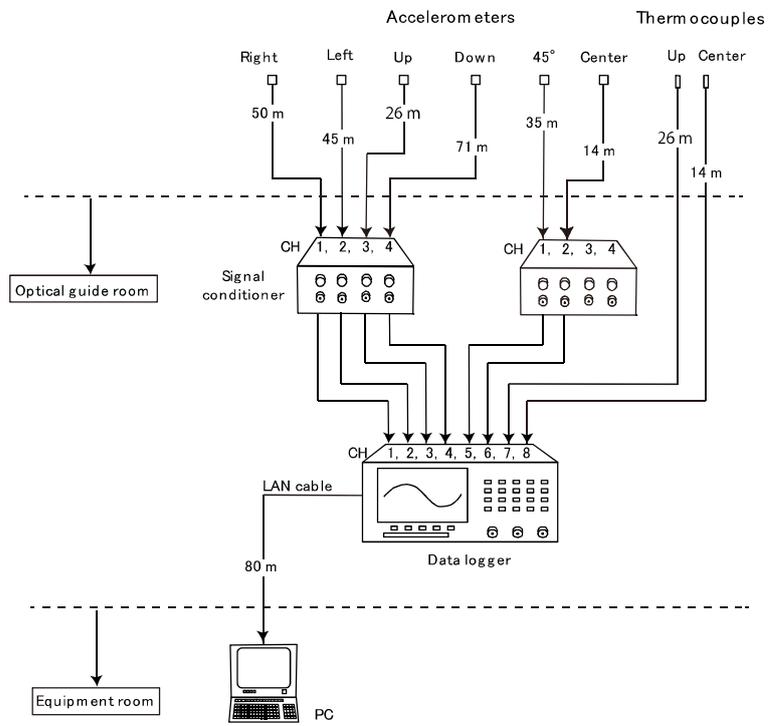

Fig. 7.  Block diagram for measuring devices.

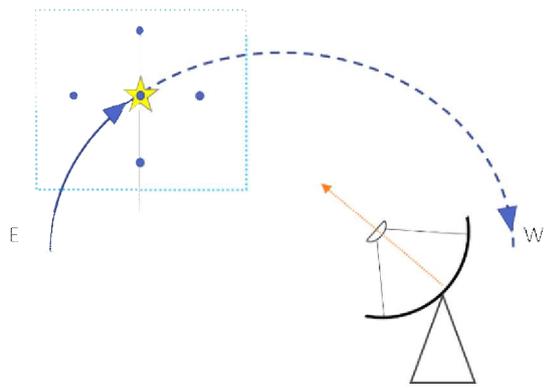
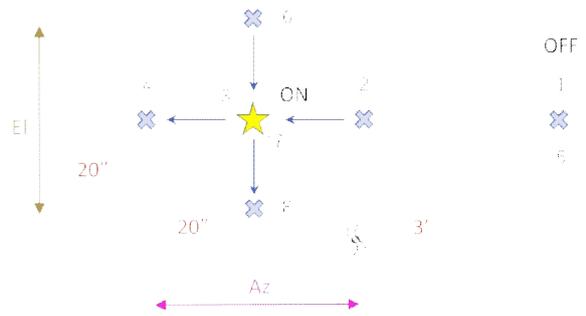

(a) Tailing targeted heavenly body.   (b) Enlargement of framed region in (a).

Fig. 8.   Overview of the pointing observation.

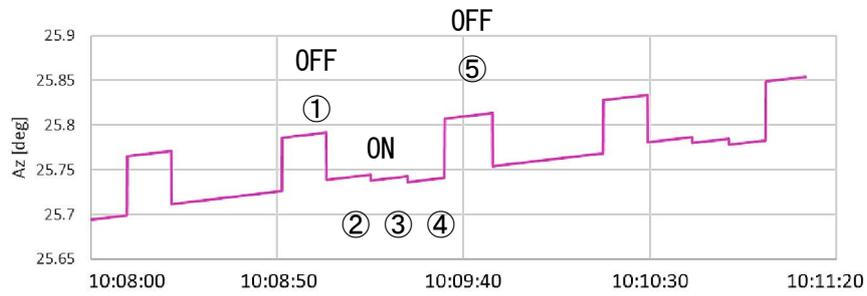

(a) Azimuth encoder data.

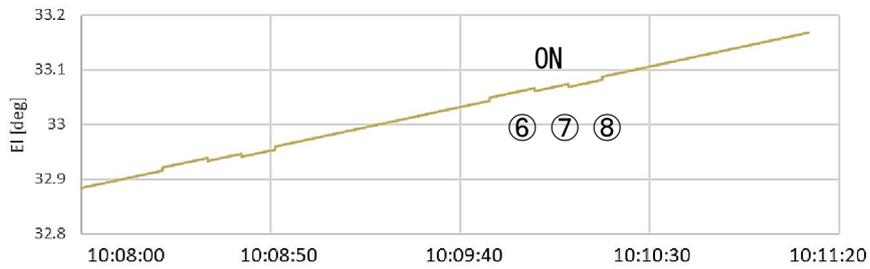

(b) Elevation encoder data.

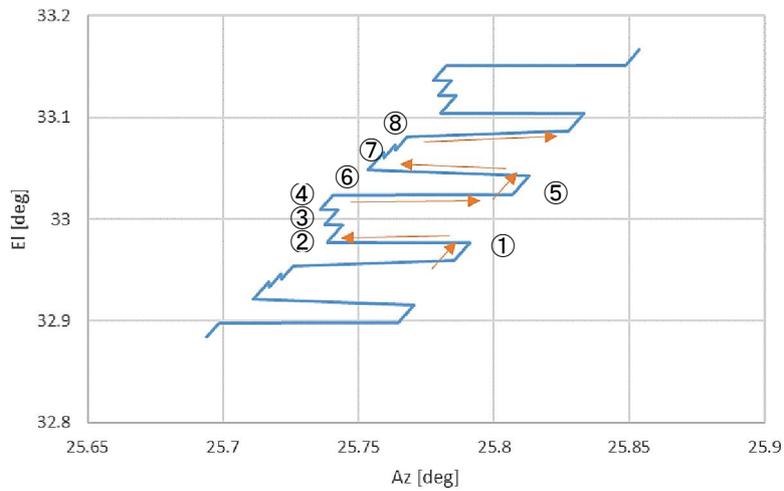

(c) Azimuth and Elevation relation.

Fig. 9.　Azimuth and Elevation encoder data at cross scan.

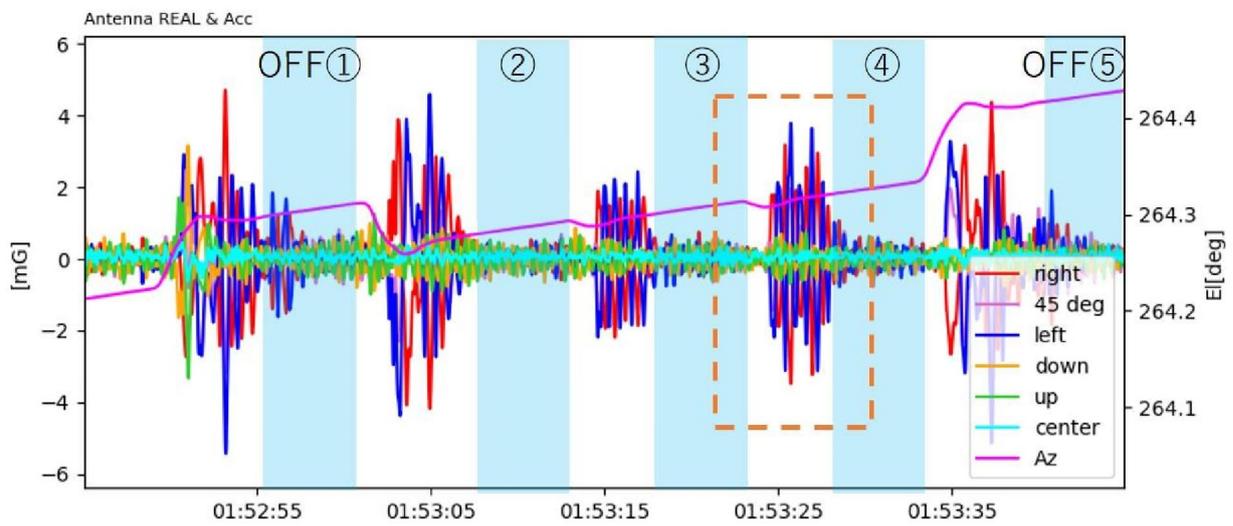

Fig. 10.   Measured acceleration and Azimuth encoder data: Scanning in Azimuth direction under no wind: $W$=0.5 m/s.

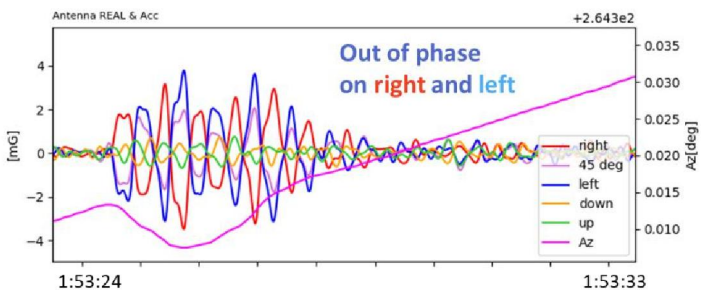

(a) Enlargement of framed region in Fig.10.

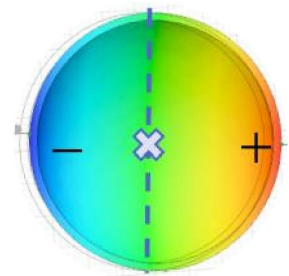

(b) Vibration mode: $N$=1.

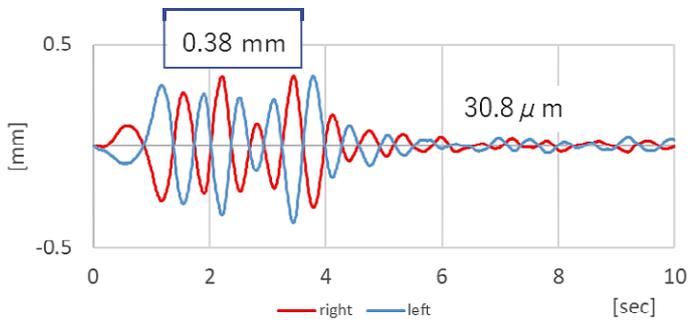

(c) Displacements: right and left positons (1:53:23.5〜)

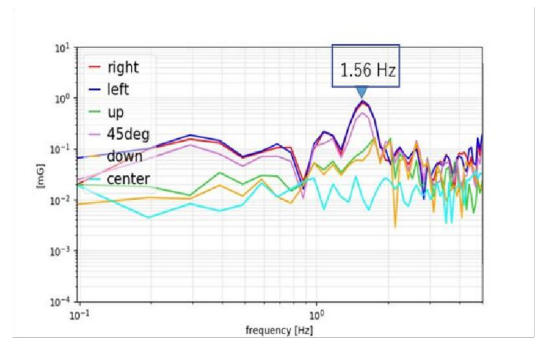

(d) Frequency components: 1:53:21~

Fig. 11.　Measured acceleration results, etc.: scanning in Azimuth direction under no wind: $W$=0.5 m/s.

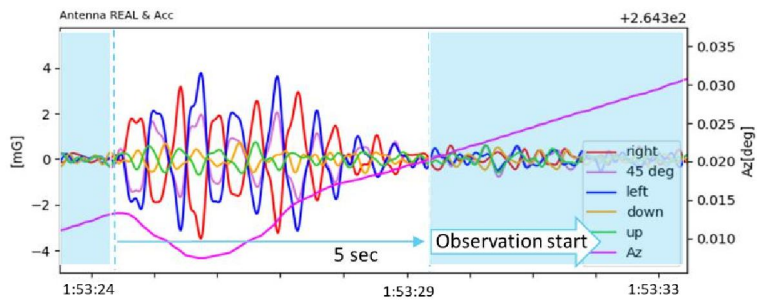

(a) Enlargement of framed region in Fig.10.

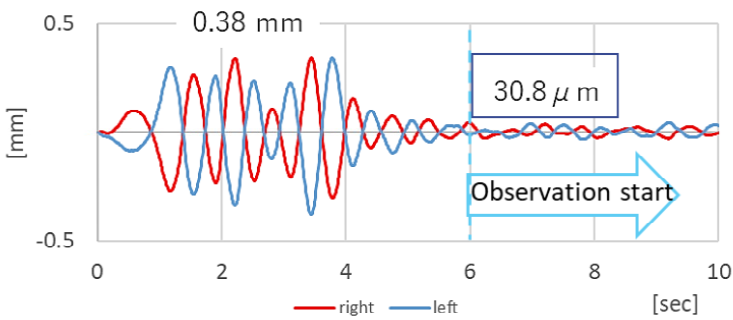

(b) Displacements: right and left positions (1:53:23.5〜).

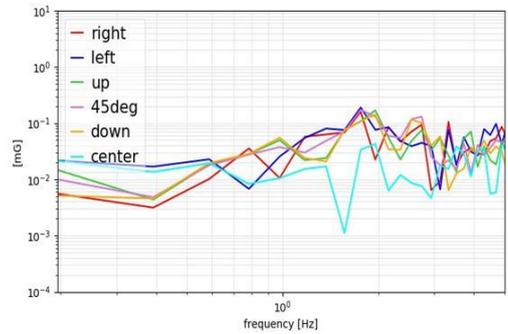

(c) Frequency components: 1:53:29~.

Fig. 12.  Measured results after observation start.

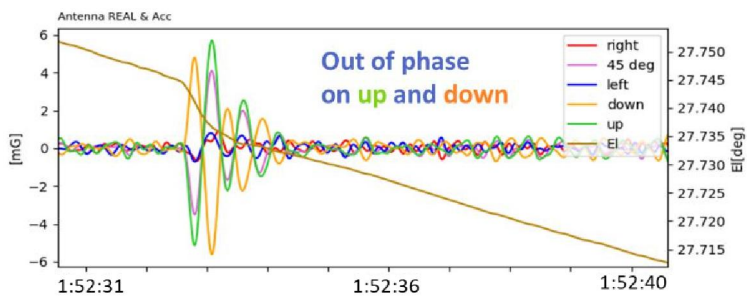 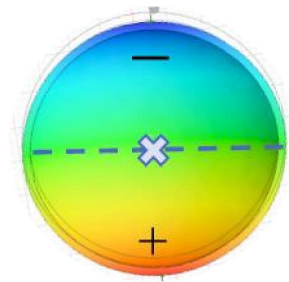

(a) Measured acceleration and Elevation encoder data.   (b) Vibration mode: $N$=1.

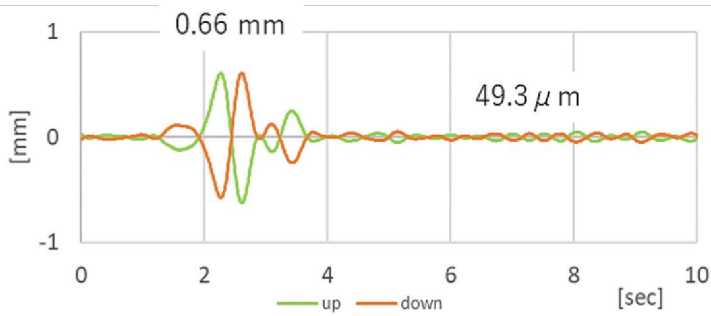 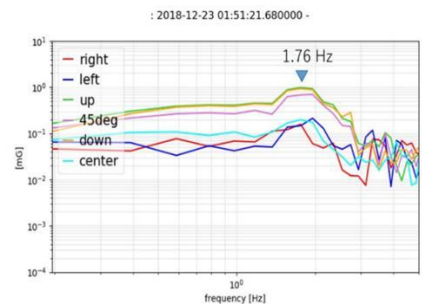

(c) Displacements: up and down positions, (1:52:30.5〜).   (d) Frequency components, (1:52:31~).

Fig. 13. Measured results: scanning in Elevation direction under no wind: $W$=0.5 m/s.

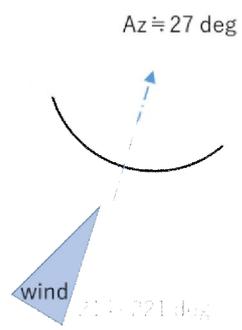

Fig. 14. Wind direction: Back side.

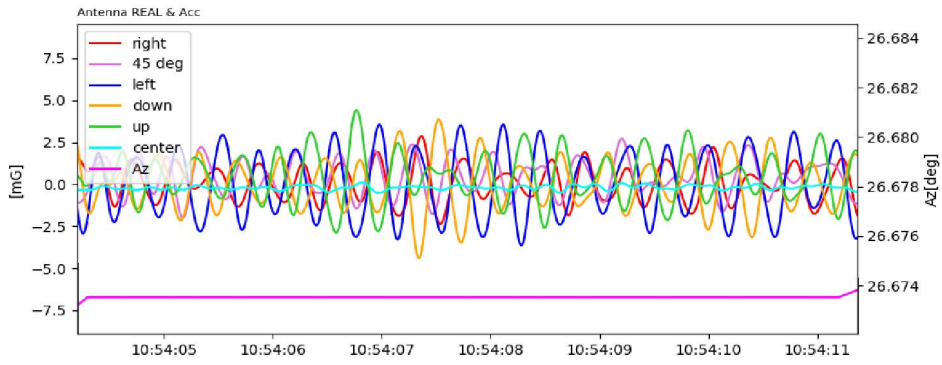

(a) Measured acceleration and Azimuth encoder data.

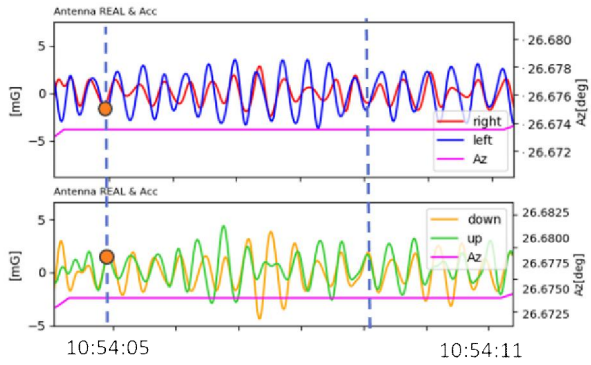

(b) Separated accelerations in (a) 10:54:05~,
Upper: right-left, lower: up-down

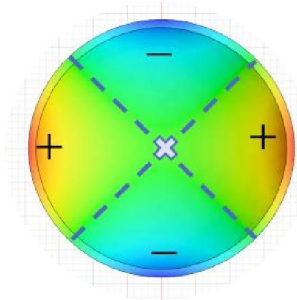

(c) Vibration mode: $N$=2.

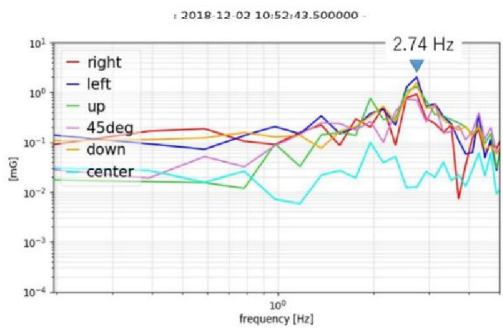

(d) Frequency components: 10:54:05~.

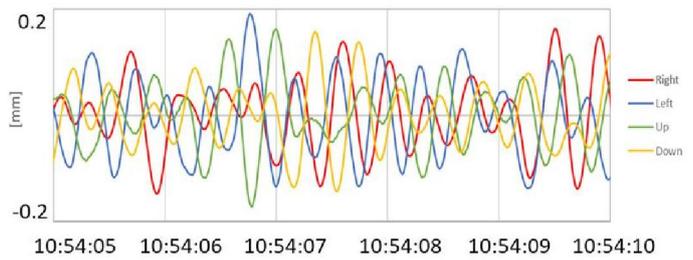

(e) Displacements.

Fig. 15. Measured results: scanning in Azimuth direction under strong wind: $El = 36°$, $W \approx 5.0$ m/s.

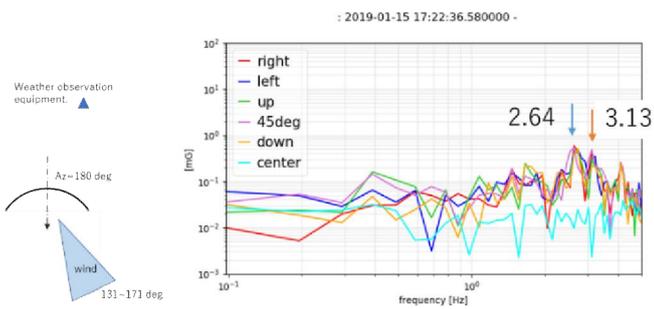

(a) Wind direction: Diagonal front side, $W \approx 5.7$ m/s.

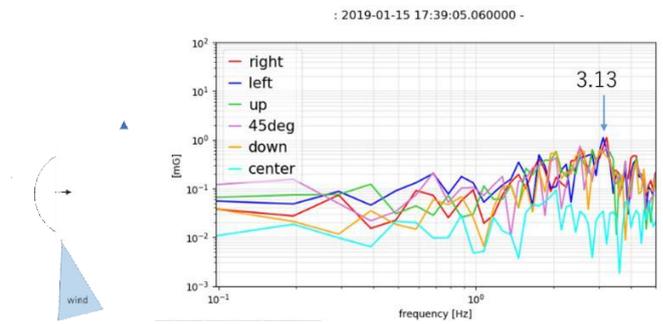

(b) Wind direction: Side, $W \approx 7.9$ m/s.

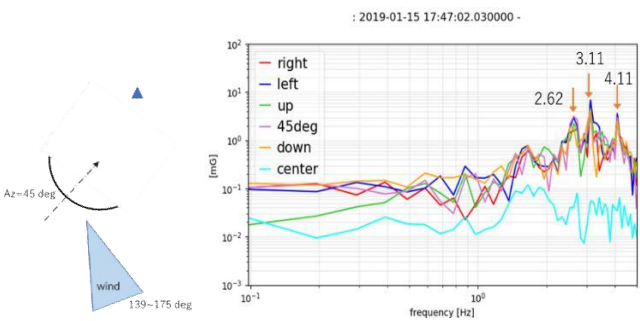

(c) Wind direction: Oblique back, $W \approx 5.7$ m/s.

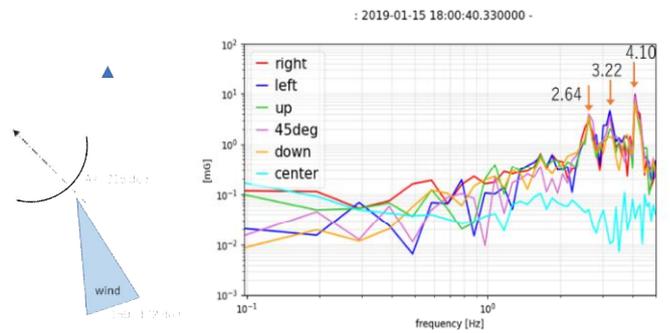

(d) Wind direction: Back side, $W \approx 13.3$ m/s.

Fig. 16. Frequency components for different wind directions, $El=20°$.

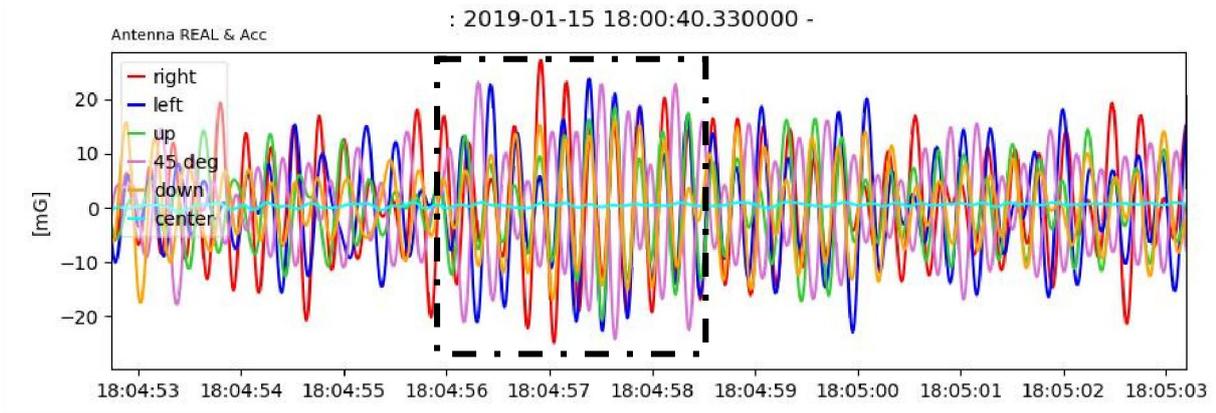

(a) Measured accelerations.

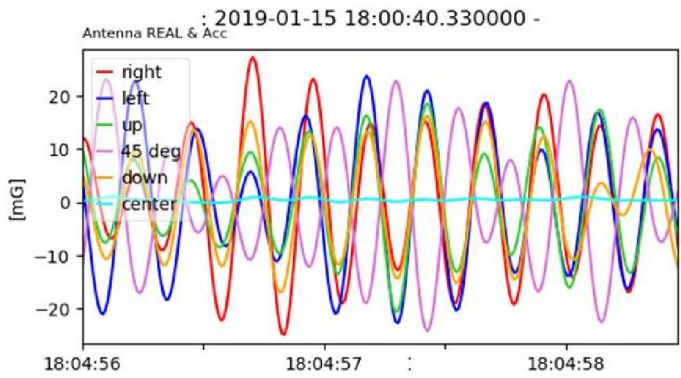

(b) Enlargement of framed region in (a).

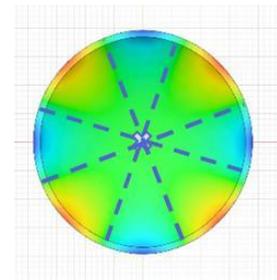

(c) Vibration mode: $N$=4.

Fig. 17. Measured results: Wind direction: Back side, $W$=13.3 m/s.

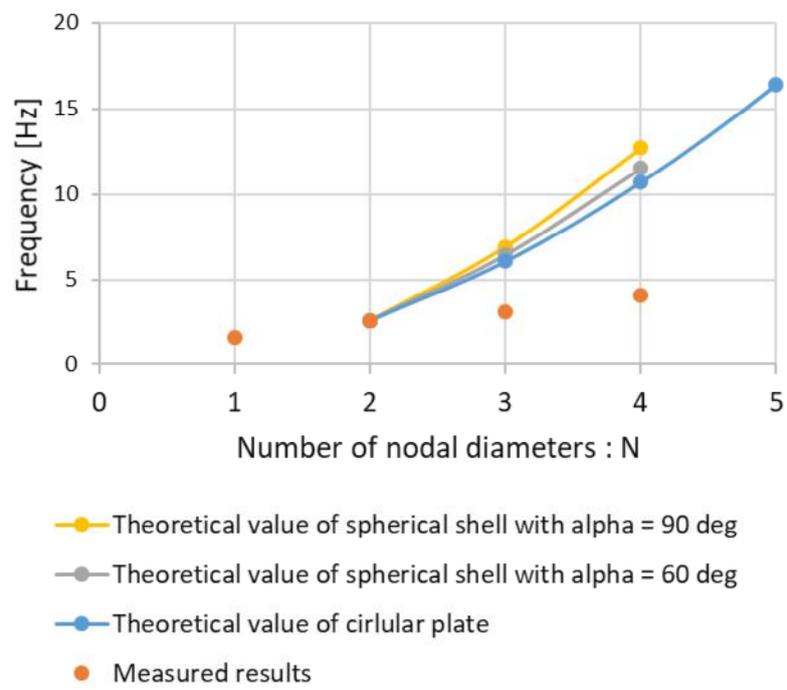

Fig. 18. Measured frequencies and theoretical values with *N* [19, 20].

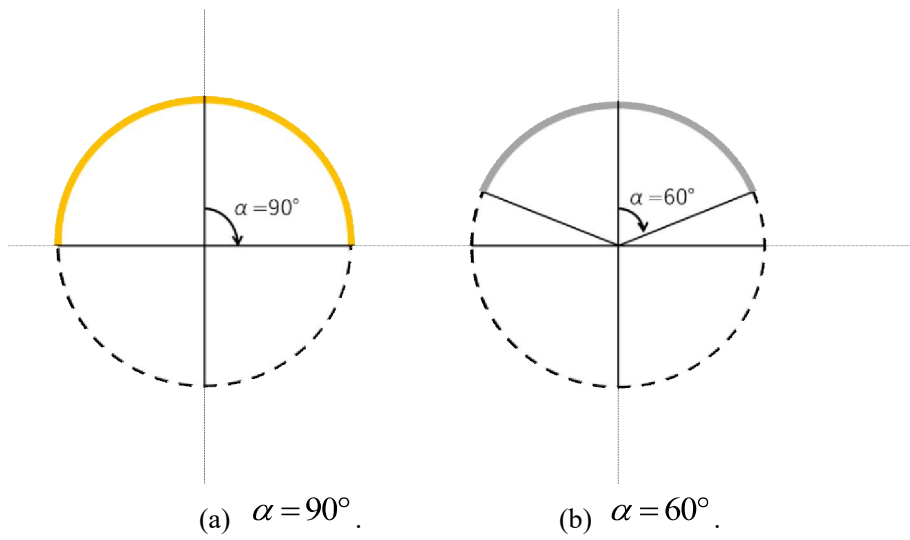

(a) $\alpha = 90°$.      (b) $\alpha = 60°$.

Fig. 19. Spherical shell with $\alpha = 90°$, $\alpha = 60°$ [20].